\documentclass[useAMS,usenatbib]{mn2e}
\usepackage{amssymb}
\usepackage{natbib}
\usepackage{times}
\usepackage{graphics,epsfig}
\usepackage{graphicx}
\usepackage{amsmath}
\usepackage{amssymb}
\usepackage{comment}
\usepackage{supertabular}
\usepackage[flushleft]{threeparttable}

\usepackage{graphicx}
\usepackage{amsmath}
\usepackage{color}

\newcommand{\lya}  {\ensuremath{{\rm Lyman}\alpha}}

\newcommand{\kms}{km~s$^{-1}$}

\newcommand{\ms}{\ensuremath{\rm M_{\odot}}}

\newcommand{\HI}{H{\sc i}}

\title[\HI~scale height in dwarf galaxies]{\HI~scale height in dwarf galaxies}
\author [N. N. Patra]{	Narendra Nath Patra$^{1}$ \thanks {E-mail: narendra@rri.res.in} \\
	$^{1}$ Raman Research Institute, C. V. Raman Avenue, Sadashivanagar, Bengaluru 560080, India\\
}
\date {}
\begin{document}
\maketitle

\begin{abstract}

Assuming a vertical hydrostatic equilibrium in the baryonic discs, joint Poisson's-Boltzmann equation was set up and solved numerically in a sample of 23 nearby dwarf galaxies from the LITTLE-THINGS survey. This is the largest sample to date for which detailed hydrostatic modeling is performed. The solutions of the Poisson's-Boltzmann equation provide a complete three-dimensional distribution of the atomic hydrogen (\HI) in these galaxies. Using these solutions, we estimate the vertical scale height (defined as the Half Width at Half Maxima (HWHM) of the density distribution) of the \HI~as a function of radius. We find that the scale height in our sample galaxies varies between a few hundred parsecs at the center to a few kiloparsecs at the edge. These values are significantly higher than what is observed in spiral galaxies. We further estimate the axial ratios to investigate the thickness of the \HI~discs in dwarf galaxies. For our sample galaxies, we find a median axial ratio to be 0.40, which is much higher than the same observed in the Milky Way. This indicates that the vertical hydrostatic equilibrium results in thicker \HI~discs in dwarf galaxies naturally.

\end{abstract}


\section{Introduction}

The local volume dwarf galaxies can be regarded as the analogs of the galaxies in the early universe. According to the hierarchical model of structure formation, the smaller galaxies form first, and then they merge to form larger galaxies. In that sense, local volume dwarf galaxies can be considered to be representative of the building blocks of the early universe. Not only that, the star formation rate and metallicities observed in these galaxies are very similar to what is expected in the Inter-Stellar Medium (ISM) of the early galaxies \citep{hunter08,ekta08,bigiel10b}. Hence, it is essential to investigate the structure and the distribution of the ISM in these galaxies. Especially, the atomic hydrogen (\HI) component, which dominates the ISM at every radius in these galaxies \citep{hunter12,schruba12}.

The \HI~in dwarf galaxies acts as the fuel reservoir for star formation and can significantly influence the evolution of such galaxies. Though, stars form out of molecular clouds, no significant molecular gas has been detected in these galaxies despite substantial efforts \citep[see, e.g.,][]{taylor98b,schruba12}. Moreover, the \HI~in these galaxies found to follow the well known Kennicutt-Schmidt law \citep{kennicutt98b,bigiel10b,roychowdhury09,roychowdhury11,roychowdhury14,patra16,roychowdhury17}, indicating a direct connection between the \HI~surface density and the star formation rate density. However, despite its tremendous importance, the three-dimensional distribution of the \HI~and its implications to the structure, shape, star formation, etc. are not well understood to date. For instance, the volume density of the \HI~and the mid-plane pressure could help in understanding the star formation in connection to its instability criteria \citep{toomre64,goldreich65}. Not only that, but it can also provide crucial inputs to the development of the theoretical frameworks for star formation processes in the ISM and origin of the star formation laws \citep{wolfire95b,dekel19}. 

For example, \citet[][]{bacchini19a} (see also, \citet{bacchini19b}) employed the hydrostatic equilibrium condition in a sample of 12 nearby disc galaxies to calculate the volume densities of the gas (atomic and molecular) and the star formation rates to derive a Volumetric Star Formation (VSF) law. They found that their VSF shows much less scatter than what is observed in the traditional Kennicutt-Schmidt law \citep{kennicutt98b} determined using surface densities. This indicates that the volume densities of different ISM phases might be more fundamental (and hence the VSF) to star formation processes than the surface densities.

Not only that, but the distribution of gas in the vertical direction has many other implications too. For example, the thickness of the \HI~discs in dwarf galaxies (which are analogs to the galaxies in the early universe) mainly decides the absorption cross-section to the background quasars, which in turn determines the average number of absorbers expected per unit redshift \citep[see, e.g.][]{zwaan05b,patra13}. This could be crucial in understanding the origin of the high-redshift Damped \lya~systems \citep[see, e.g., ][]{wolfe86,prochaska97b,haehnelt98a}. The thickness of the \HI~discs further decides the sizes of the bubbles/super-bubbles up to which it can grow before breaking out into the Inter-Galactic Medium (IGM). In that sense, it controls the mixing of the high metallicity gas into the IGM. Moreover, the intrinsic shape of the galaxies is a fundamental parameter that must be produced by the galaxy evolutionary models to be consistent with the observations.

Hence, in these respects, it is imperative to estimate the distribution of the \HI~and, subsequently, the scale heights (defined as Half-Width at Half Maxima) in galaxies. However, measuring the \HI~distribution directly from observation is not straight forward. Even for the Milky Way, having to reside within it, rigorous modeling is required to estimate the three-dimensional distribution of the \HI~from the observed brightness temperature profiles \citep[see, e.g.,][]{kalberla07,kalberla08}. For external galaxies, on the other hand, the lack of spatial resolution and the line-of-sight projection effects make it more challenging to estimate the volume densities using direct measurements. Only for special cases, e.g., for a constant scale height as a function of radius observed in an edge-on orientation, allows one to measure the scale height directly. However, for real galaxies, the scale height is never found to be constant, and hence, effectively, in all cases, detailed modeling assisted by observation is needed \citep[see, e.g., ][]{narayan02a,narayan02b,banerjee07,patra14}.

In this work, we model the discs of dwarf galaxies as a two-component system consists of stars and \HI~in hydrostatic equilibrium under their mutual gravity in the external force field of the dark matter halo. Under this assumption, we set up the joint Poisson's-Boltzmann equation of hydrostatic equilibrium and numerically solve it. The solution of the Poisson's-Boltzmann equation provides the density of the gas and stars as a function of the height from the mid-plane ($z$) at different radii. This, in turn, provides a detailed three-dimensional distribution of the \HI~in a galaxy. A number of previous studies used similar modeling techniques to estimate the gas/stellar distribution in galaxies theoretically. For example, in early studies, \citet{narayan02b} solved the Poisson's-Boltzmann equation for a three-component disc in a self-consistent manner. Later, several other studies used this method to estimate the \HI/molecular scale heights in spiral galaxies \citep{banerjee07,banerjee08,banerjee10,patra18a,patra19b}. \citet{banerjee11b} used this technique to solve the hydrostatic equilibrium equation in four dwarf galaxies. They found that the \HI~scale height in dwarf galaxies are larger than what is found in spiral galaxies. In this paper, we solve the joint Poisson's-Boltzmann equation of hydrostatic equilibrium self-consistently in a sample of 23 galaxies from the Local Irregulars That Trace Luminosity Extremes - The \HI~Nearby Galaxy Survey (LITTLE-THINGS) \citep{hunter12}. This is the largest sample to date for which a theoretical determination of the three-dimensional density distribution of the \HI~is attempted.


\section{Sample}

We choose our sample galaxies from the LITTLE-THINGS survey for which all the necessary data are available publicly. The LITTLE-THINGS sample consists of 41 galaxies, which are observed in \HI~using the Very Large Array (VLA). Out of these 41 galaxies, 37 are classified as dwarf irregulars, whereas four galaxies are classified as Blue Compact Dwarfs (BCDs). As we will see in the next section, the rotation curve of a galaxy plays a crucial input to the hydrostatic equation. Hence, it is necessary to have the rotation curve measured for a galaxy to consider it for hydrostatic modeling. Unfortunately, out of 41 sample galaxies in the LITTLE-THINGS survey, the rotation curve could be extracted only for 26 galaxies \cite[see][for more details]{oh15}. Moreover, to avoid any divergence while solving the Poisson's-Boltzmann equation, we needed to fit the rotation curves of our sample galaxies with a smooth function (see \S 3.1 for more details). Traditionally, a Brandt profile \citep{brandt60} is used to represent a typical rotation curve. For three galaxies, i.e., DDO 46, F564-V3, and IC 1613, the rotation curves could not be well represented by a Brandt profile (or a more simpler straight line) due to their sharply declining nature at outer radii. We exclude these three galaxies from our final sample. This leads to a total of 23 dwarf irregular galaxies for which we set up the Poisson's-Boltzmann equation and solve numerically to obtain the three-dimensional distribution of the \HI~in them. In table~\ref{tab:samp} we present the basic properties of our sample galaxies. As can be seen from the table, our sample galaxies trace the low mass end of the \HI~mass function having \HI~masses $<10^9$ \ms. These low mass galaxies are particularly interesting due to their similarities to the galaxies expected in the early universe.

\begin{table*}
\begin{threeparttable}
\begin{tabular}{lcccccccc}
\hline
Name & RA (J2000) & DEC (J2000)  & Dist &  $\rm M_V$  & $\rm R_H$  & $\rm R_D$ & $\rm \log M_{HI}$ & i \\
     & (h m s)    & ($^o$ $^\prime$ $^{\prime \prime}$) & (Mpc) & (mag) & ($^\prime$) & (kpc) & ($\rm M_{\odot}$) & ($^o$) \\
(1)   &   (2)      & (3)                                 & (4)   & (5)   &   (6)       &  (7)  &      (8)          &  (9) \\
\hline
CVnIdwA & 12 38 39.2 & +32 45 41.0 & 3.6 & -12.4 & 0.87 & 0.41 & 7.67 & 66.5  \\ 
DDO 101 & 11 55 39.1 & +31 31 9.9 & 6.4 & -15.0 & 1.05 & 0.94 & 7.36 & 51.0  \\ 
DDO 126 & 12 27 06.6 & +37 08 15.9 & 4.9 & -14.9 & 1.76 & 0.87 & 8.16 & 65.0  \\ 
DDO 133 & 12 32 55.2 & +31 32 19.1 & 3.5 & -14.8 & 2.33 & 1.24 & 8.02 & 43.4  \\ 
DDO 154 & 12 54 05.7 & +27 09 09.9 & 3.7 & -14.2 & 1.55 & 0.59 & 8.46 & 68.2  \\ 
DDO 168 & 13 14 27.3 & +45 55 37.3 & 4.3 & -15.7 & 2.32 & 0.82 & 8.47 & 46.5  \\ 
DDO 210 & 20 46 51.6 & -12 50 57.7 & 0.9 & -10.9 & 1.31 & 0.17 & 6.30 & 66.7  \\ 
DDO 216 & 23 28 34.7 & +14 44 56.2 & 1.1 & -13.7 & 4.00 & 0.54 & 6.75 & 63.7  \\ 
DDO 43 & 07 28 17.7 & +40 46 08.3 & 7.8 & -15.1 & 0.89 & 0.41 & 8.23 & 40.6  \\ 
DDO 47 & 07 41 55.3 & +16 48 07.1 & 5.2 & -15.5 & 2.24 & 1.37 & 8.59 & 45.5  \\ 
DDO 50 & 08 19 03.7 & +70 43 24.6 & 3.4 & -16.6 & 3.97 & 1.10 & 8.85 & 49.7  \\ 
DDO 52 & 08 28 28.4 & +41 51 26.5 & 10.3 & -15.4 & 1.08 & 1.30 & 8.43 & 43.0  \\ 
DDO 53 & 08 34 06.4 & +66 10 47.9 & 3.6 & -13.8 & 1.37 & 0.72 & 7.72 & 27.0  \\ 
DDO 70 & 10 00 00.9 & +05 20 12.9 & 1.3 & -14.1 & 3.71 & 0.48 & 7.61 & 50.0  \\ 
DDO 87 & 10 49 34.9 & +65 31 47.9 & 7.7 & -15.0 & 1.15 & 1.31 & 8.39 & 55.5  \\ 
Haro 29 & 12 26 18.4 & +48 29 40.4 & 5.9 & -14.6 & 0.84 & 0.29 & 7.80 & 61.2  \\ 
Haro 36 & 12 46 56.6 & +51 36 47.3 & 9.3 & -15.9 & - & 0.69 & 8.16 & 70.0  \\ 
IC 10 & 00 20 18.9 & +59 17 49.9 & 0.7 & -16.3 & - & 0.40 & 7.78 & 47.0  \\ 
NGC 1569 & 04 30 46.2 & +64 51 10.3 & 3.4 & -18.2 & - & 0.38 & 8.39 & 69.1  \\ 
NGC 2366 & 07 28 53.4 & +69 12 49.6 & 3.4 & -16.8 & 4.72 & 1.36 & 8.84 & 63.0  \\ 
NGC 3738 & 11 35 46.9 & +54 31 44.8 & 4.9 & -17.1 & 2.40 & 0.78 & 8.06 & 22.6  \\ 
UGC 8508 & 13 30 44.9 & +54 54 32.4 & 2.6 & -13.6 & 1.28 & 0.27 & 7.28 & 82.5  \\ 
WLM & 00 01 59.9 & -15 27 57.2 & 1.0 & -14.4 & 5.81 & 0.57 & 7.85 & 74.0  \\ 
\hline

\end{tabular}
\label{tab:samp}
\end{threeparttable}
\caption{Basic properties of our sample galaxies. The first column shows the name of the galaxies. Column (2) and (3) represent the RA and DEC coordinates, whereas column (4) denotes the distances to the galaxies. Column (5) shows the absolute $V$ band magnitude. In column (6) and (7), we quote the Holmberg radius and the disc scale length in the optical band, respectively. Column (8) and (9) show the \HI~mass and the inclination of the \HI~discs, respectively. All the data in column (4) to (8) are obtained from \citet{walter08}, whereas data in column (9) is taken from \citet{oh15}.}
\end{table*}

\section{Modelling the galaxy discs}

We consider the galactic discs in dwarf galaxies to be a two-component system consisting of an \HI~and a stellar disc. We note that no considerable amount of molecular gas has been detected in dwarf galaxies despite several significant efforts \citep{taylor98b,schruba12}. Hence, we ignore the contribution of the molecular gas in the formulation of the hydrostatic equation. Unlike spiral galaxies where the stellar discs dominate the gravity within the optical radii, in dwarf galaxies, the \HI~contributes to the disc surface density equally or more as compared to the stellar disc (over all radii). In other words, the gas discs in dwarf galaxies are more critical in deciding the hydrostatic equilibrium than the stellar discs. 

We assume that the stellar and the gas discs in dwarf galaxies are in vertical hydrostatic equilibrium under the total gravity of the baryonic discs (star+\HI) and the dark matter halo, balanced by the vertical pressure generated by the velocity dispersions of the individual components. To simplify the modeling, we further assume that both the baryonic discs are co-planar and concentric with the centers coinciding with the center of the dark matter halo. Under these assumptions, one then can write the Poisson's equation of hydrostatic equilibrium in cylindrical coordinate as 

\begin{equation}
\frac{1}{R} \frac{\partial }{\partial R} \left( R \frac{\partial \Phi_{tot}}{\partial R} \right) + \frac{\partial^2 \Phi_{tot}}{\partial z^2} = 4 \pi G \left( \sum_{i=1}^{2} \rho_{i} + \rho_{h} \right)
\label{eq1}
\end{equation}

\noindent where $\Phi_{tot}$ is the total gravitational potential due to the baryonic discs and the dark matter halo. Whereas, $\rho_i$ denotes the volume density of the two disc components (stars and \HI). $\rho_h$ represents the volume density of the underlying dark matter halo, which is an input parameter to our modeling and estimated by the mass-modeling of the rotation curve.

In Eq.~\ref{eq1}, $\Phi_{tot}$ is not a directly measurable quantity. Instead, it relates to the pressure gradient as a requirement of the static equilibrium

\begin{equation}
\frac{\partial }{\partial z} \left(\rho_i {\langle {\sigma}_z^2 \rangle}_i \right) + \rho_i \frac{\partial \Phi_{tot}}{\partial z} = 0
\label{eq2}
\end{equation}

\noindent i.e., in static equilibrium, the gradient of the potential would be balanced by the gradient in pressure.

Using Eq.~\ref{eq2}, we can rewrite Eq.~\ref{eq1} as

\begin{equation}
\begin{split}
{\langle {\sigma}_z^2 \rangle}_i \frac{\partial}{\partial z} \left( \frac{1}{\rho_i} \frac{\partial \rho_i}{\partial z} \right) &= \\ 
&-4 \pi G \left( \rho_s + \rho_{HI} + \rho_h \right)\\ 
&+ \frac{1}{R} \frac{\partial}{\partial R} \left( R \frac{\partial \Phi_{tot}}{\partial R} \right)
\end{split}
\label{eq3}
\end{equation}

\noindent where $\rho_s$, $\rho_{HI}$ are the volume density of stars and \HI~respectively. The term on the left-hand side describes the vertical pressure, which is decided by the observed velocity dispersion of an individual component. It should be mentioned here that, for simplicity, we do not consider any variation of $\sigma_i$ along $z$. The first term on the RHS is the gravity term, which is coupled to all the three entities, i.e., stars, \HI~and dark matter. The second term on the RHS is an outcome of the radial gradient of the potential, which can be estimated using the observed rotation curve of the galaxy.

\begin{equation}
{\left( R \frac{\partial \Phi_{total}}{\partial R} \right)}_{R,z} = {(v_{rot}^2)}_{R,z}
\label{eq4}
\end{equation}

Assuming that the rotation curve does not change as a function of height ($z$), Eq.~\ref{eq3} can be simplified to 

\begin{equation}
\begin{split}
{\langle {\sigma}_z^2 \rangle}_i \frac{\partial}{\partial z} \left( \frac{1}{\rho_i} \frac{\partial \rho_i}{\partial z} \right) &= \\
&-4 \pi G \left( \rho_s + \rho_{HI} + \rho_h \right)\\ 
&+ \frac{1}{R} \frac{\partial}{\partial R} \left( v_{rot}^2 \right)
\end{split}
\label{eq5}
\end{equation}

Eq.~\ref{eq5} represents two coupled second-order ordinary partial differential equations in $\rho_{HI} (z)$ and $\rho_{s} (z)$. These equations must be solved in all radii to obtain a total three-dimensional distribution of \HI. To solve Eq.~\ref{eq5}, one needs several input parameters which are described in the next subsection.

\subsection{Input parameters}
\label{inputs}

\begin{figure*}
\begin{center}
\begin{tabular}{c}
\resizebox{1.\textwidth}{!}{\includegraphics{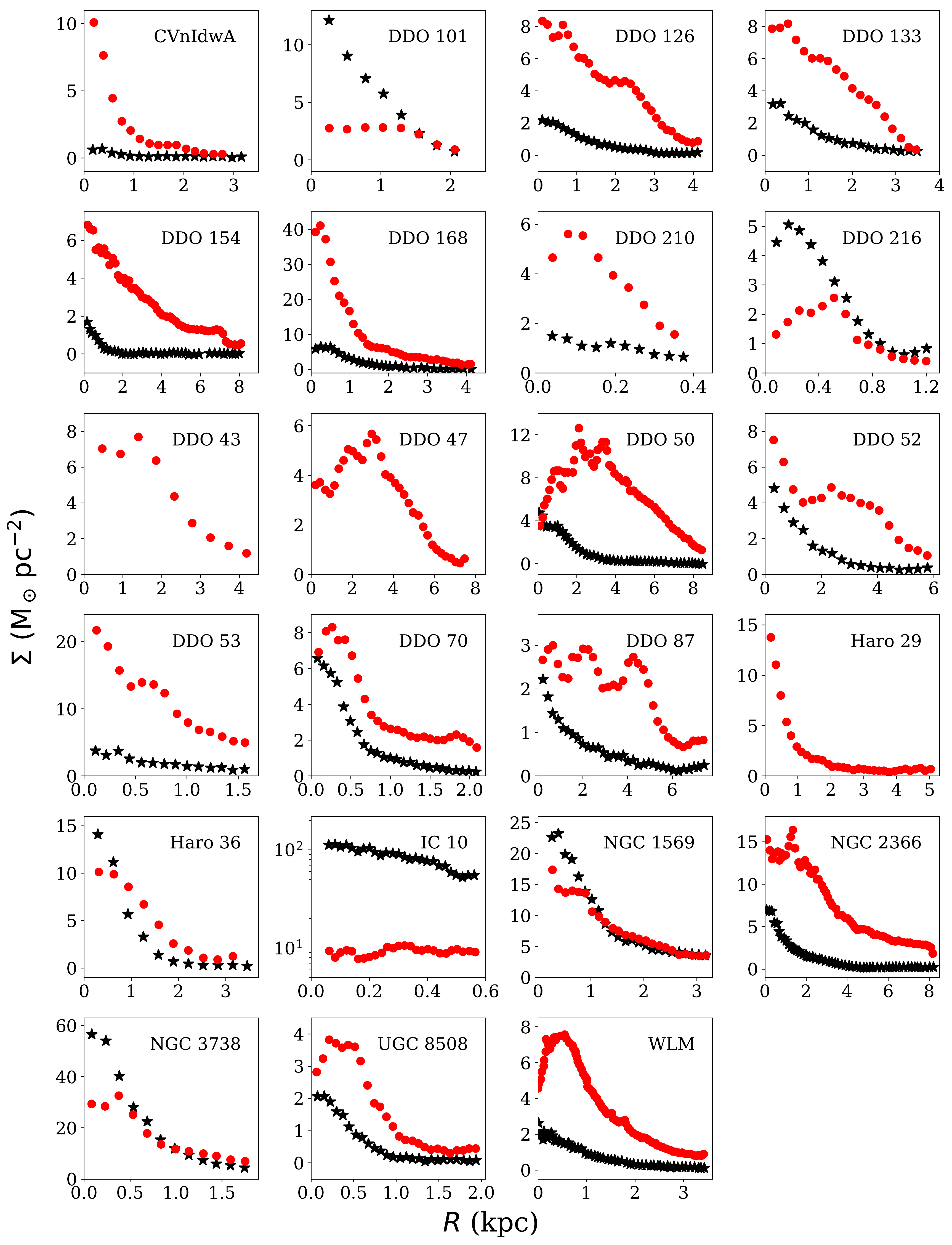}}
\end{tabular}
\end{center}
\caption{Surface density profiles of our sample galaxies. Individual panels represent the surface density profiles of different galaxies, as quoted in the top right corners of every panel. The black filled asterisks represent the stellar surface density profiles, whereas the red filled circles represent the atomic gas surface density profiles. The data are compiled from \citet{oh15}.}
\label{sden}
\end{figure*}

There are four primary inputs to Eq.~\ref{eq5} which are 1) the surface densities of individual disc components which together with the dark matter halo provides the gravity (first term on the RHS) 2) The velocity dispersions of individual components, $\sigma_{z,i}$ which provides the pressure to hold the gravity in the vertical direction (first term on LHS), 3) the rotation curve which is necessary to compute the radial term (second term in RHS) and 4) the dark matter halo parameters. These input parameters are essential to solve the hydrostatic equilibrium equation. 

The surface densities are one of the critical input parameters which provide gravity and primarily decide the stable equilibrium condition. We adopt the \HI~and the stellar surface density profiles of our sample galaxies from \citet{oh15}. \citet{oh15} produced the \HI~surface density profiles using the robust weighted spectral cubes from the LITTLE-THINGS survey after applying tilted ring geometric parameters. This results in \HI~surface density profiles with a higher degree of accuracy than a simple azimuthal averaging assuming a flat disc \citep[see][for more details]{oh15}.

The stellar surface density profiles are produced by \citet{oh15} using $Spitzer$ IRAC 3.6 $\mu m$ maps from the SINGS (The Spitzer Infrared Nearby Galaxies Survey) \citep{kennicutt03} and the LVL (Local Volume Legacy) survey \citep{dale09} data. As compared to the optical images, the 3.6 $\mu m$ maps are less affected by dust and hence, provide a more accurate estimate of the stellar mass. Not only that, but the 3.6 $\mu m$ band is also insensitive to the radiation from young stellar populations, which is responsible for a large amount of radiation but constitutes only a small fraction of the total mass. The 3.6 $\mu m$ images are useful to trace the old stellar population dominant in the late-type dwarf galaxies. Thus using 3.6 $\mu m$ maps provides a robust estimate of the stellar mass surface density profile. The mass to light ratio at 3.6 $\mu m$, $\Upsilon_*^{3.6}$ is calculated by using the empirical formula based on the stellar population synthesis model of \citet{bruzual03} and \citet{bell01}. They find a typical value of $\Upsilon_*^{3.6} \sim 0.35$ for their sample galaxies \citep[see][for more details]{oh11b}.

Thus generated \HI~and stellar surface density profiles of our sample galaxies are shown in Fig.~\ref{sden}. As can be seen from the figure, for most of the galaxies, the \HI~surface density is comparable or larger than the stellar surface density. This implies that for our sample galaxies, \HI~dominates the surface density at almost all radii. We note here that the \HI~surface densities shown in Fig.~\ref{sden} (and used to solve the hydrostatic equilibrium equation) is corrected by a factor 1.4 to account for Helium and metals present in the ISM. For three galaxies, i.e., DDO 43, DDO 47, and Haro 29, the stellar surface density profiles could not be produced due to the unavailability of the IRAC 3.6 $\mu m$ data \citep[see][for more details]{oh15}. For these galaxies, we assume that the \HI~dominates the baryonic disc and solve Eq.~\ref{eq5} by putting $\rho_s=0$. In fact, \citet{oh15} used gas surface density profiles only to produce the mass models of these three galaxies.

The vertical velocity dispersion of individual components is another input parameter that influences the vertical \HI~scale height directly. Unlike the surface densities, the velocity dispersion determines the vertical pressure alone and hence plays a vital role in determining the solutions of Eq.~\ref{eq5}. Hence, a precise measurement of the same is necessary. However, it is found that while solving the hydrostatic equilibrium equation, stellar velocity dispersion does not modify the \HI~scale height considerably \citep[see, e.g.,][]{banerjee11b}. Moreover, a direct measurement of the stellar velocity dispersions in galaxies is tough using the existing observing methods. Considering these facts, we theoretically calculate the stellar velocity dispersion in our sample galaxies, assuming a prevailing hydrostatic equilibrium, as given in \citet{leroy08}. For this calculation, the stellar disc is assumed to have an exponential vertical density distribution with a constant scale height, $h_*$ (i.e., no flaring). The stellar scale height is then estimated using the observed flattening ratio in the optical discs of galaxies as $l*/h* = 7.3 \pm 2.2$ \citep{kregel02}, where $l_*$ is the exponential disc scale length. Assuming an isothermal stellar disc in hydrostatic equilibrium and a vertical to radial velocity dispersion ratio, $\sigma_{*,z}/\sigma_{*,r} = 0.6$, the vertical stellar velocity dispersion at any radius is then computed as $\sigma_{*,z} = 1.879 \sqrt{l_* \Sigma_*}$ where $\Sigma_*$ is the stellar surface density \citep[see Appendix-B.3 of][for more details]{leroy08}. However, recent studies using IFU data suggest that the assumption of an isothermal stellar disc always overestimates the true velocity dispersion within the disc scale length and underestimates the same beyond the disc scale length \citep[][]{mogotsi18}. Nonetheless, a theoretical approximation to the stellar velocity dispersion is justified here as it marginally influences the vertical \HI~distribution.

Unlike the stellar velocity dispersion, the \HI~velocity dispersion can be estimated easily using spectral line observations. Early low-resolution observations of the \HI~discs in external galaxies resulted in a velocity dispersion in the range between 6-13 \kms~\citep{shostak84,vanderkruit84,kamphuis93}. However, with the advent of modern-day radio telescopes, the velocity dispersions of the \HI~discs in nearby galaxies can be measured with high spectral and spatial resolutions. The \HI~spectral cubes can be used to compute a map of the second moment (MOMNT2) of the line-of-sight \HI~spectra. Very often, this MOMNT2 is used as a measure of the velocity dispersion in the \HI~disc. For example, \citet{tamburro09} used the data from the THINGS survey to estimate the velocity dispersion in spiral galaxies using the MOMNT2 maps. However, though the MOMNT2 can be used as an easily available measure of the velocity dispersion, it is known to be erroneous at low SNR regions. At low SNR, MOMNT2 only samples the peak of the \HI~spectra, which in turn artificially reduces the \HI~velocity dispersion.

To avoid this SNR problem, the \HI~spectra within a radial bin could be stacked together to produce a stacked spectrum. This stacked spectrum has a much higher SNR than any of the individual \HI~spectrum in the cube and can be used to estimate the $\sigma_{HI}$, which is now free from any SNR problem. In fact, a number of studies used this stacking technique to produce and study high SNR \HI~spectra in galaxies. For example, \citet{ianjamasimanana12} stacked \HI~spectra of THINGS galaxies to find an \HI~velocity dispersion between 10-13 \kms. \citet{stilp13} has used the data from the Very Large Array ACS Nearby Galaxy Survey Treasury Program (`VLA-ANGST'; \citet{ott12}) to stack the \HI~spectra of dwarf galaxies and found a central velocity dispersion of $\sim 5-15$ \kms. These studies repeatedly pointed out that a stacking approach results in a better estimation of the velocity dispersion in \HI~discs.

Adopting a similar approach, we stack the line-of-sight \HI~spectra in annular rings of our sample galaxies using the \HI~spectral cubes from the LITTLE-THINGS survey. We use the same method for stacking, as described in \citet{deblok08}. We first fit line-of-sight \HI~spectra with Gaussian Hermite Polynomials of order three to locate their centroids. After determining the centroids of all such spectra, we shift them to a common velocity. This, in turn, aligns the centers of all spectra, after which we stack them together to produce a high SNR stacked spectrum. We do this exercise for each annular ring in a galaxy, representing a particular radius. We choose the widths of the rings to be the same as the observed beamwidth. Thus obtained stacked spectra are then fitted with single Gaussian functions to estimate the $\sigma_{HI}$. We emphasize here that, a minimum SNR of 5 required to fit the \HI~spectra with the Gaussian Hermite polynomial \citep[see, e.g.,][]{deblok08}. Hence, we stack only those spectra in the cube which satisfy this minimum SNR criterion.

\begin{figure}
\begin{center}
\begin{tabular}{c}
\resizebox{0.48\textwidth}{!}{\includegraphics{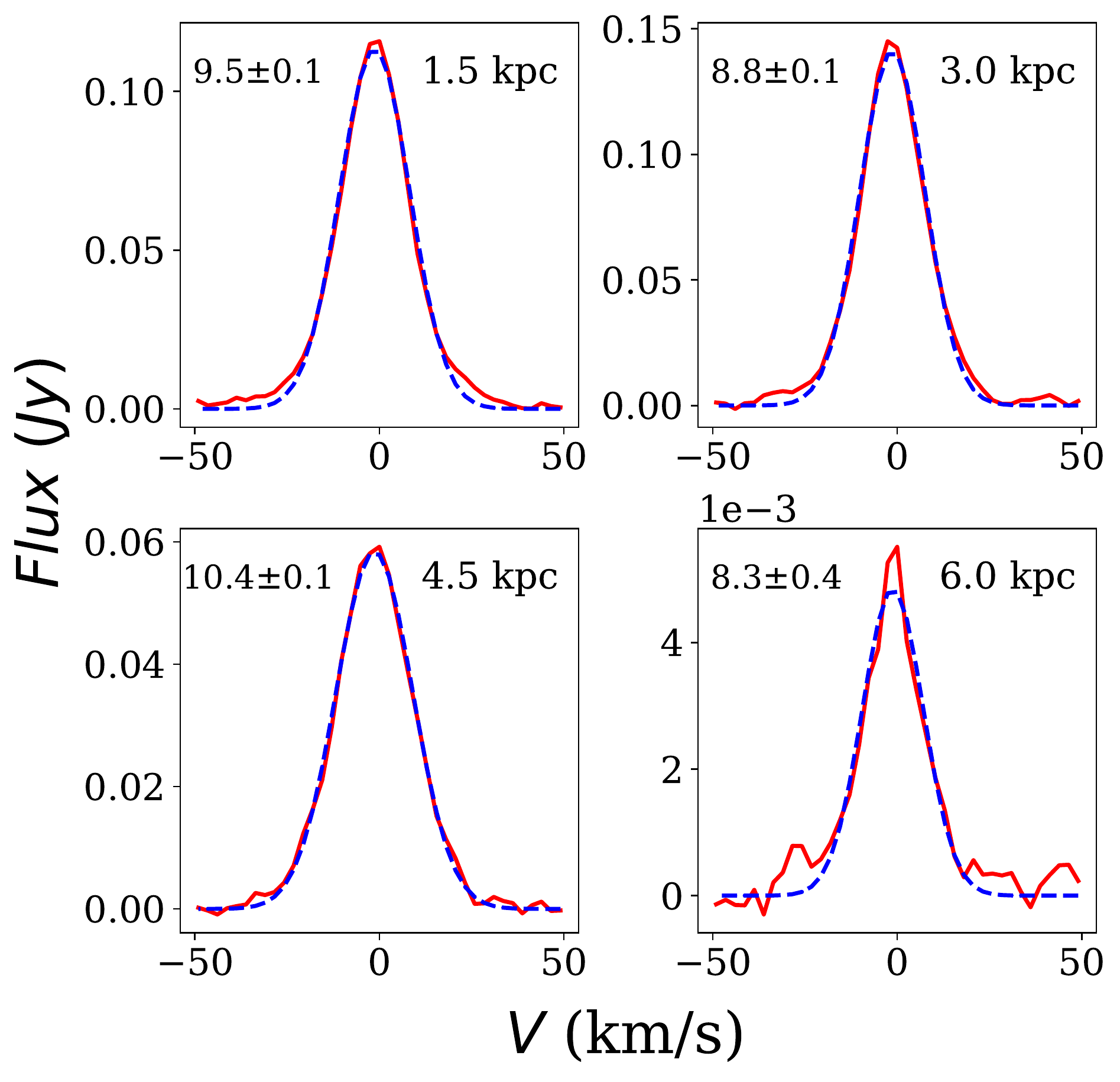}}
\end{tabular}
\end{center}
\caption{The stacked \HI~spectra and their fits to single-Gaussian profiles for DDO 154. Each panel indicates the stacked spectrum at a different radius, as quoted at the top right corner. The solid red lines indicate the stacked spectra, whereas the blue dashed lines represent a single-Gaussian fit to the same. The respective $\sigma_{HI}$ are quoted in the top left corners of the individual panels in the units of \kms.}
\label{sprof}
\end{figure}

\begin{figure}
\begin{center}
\begin{tabular}{c}
\resizebox{0.48\textwidth}{!}{\includegraphics{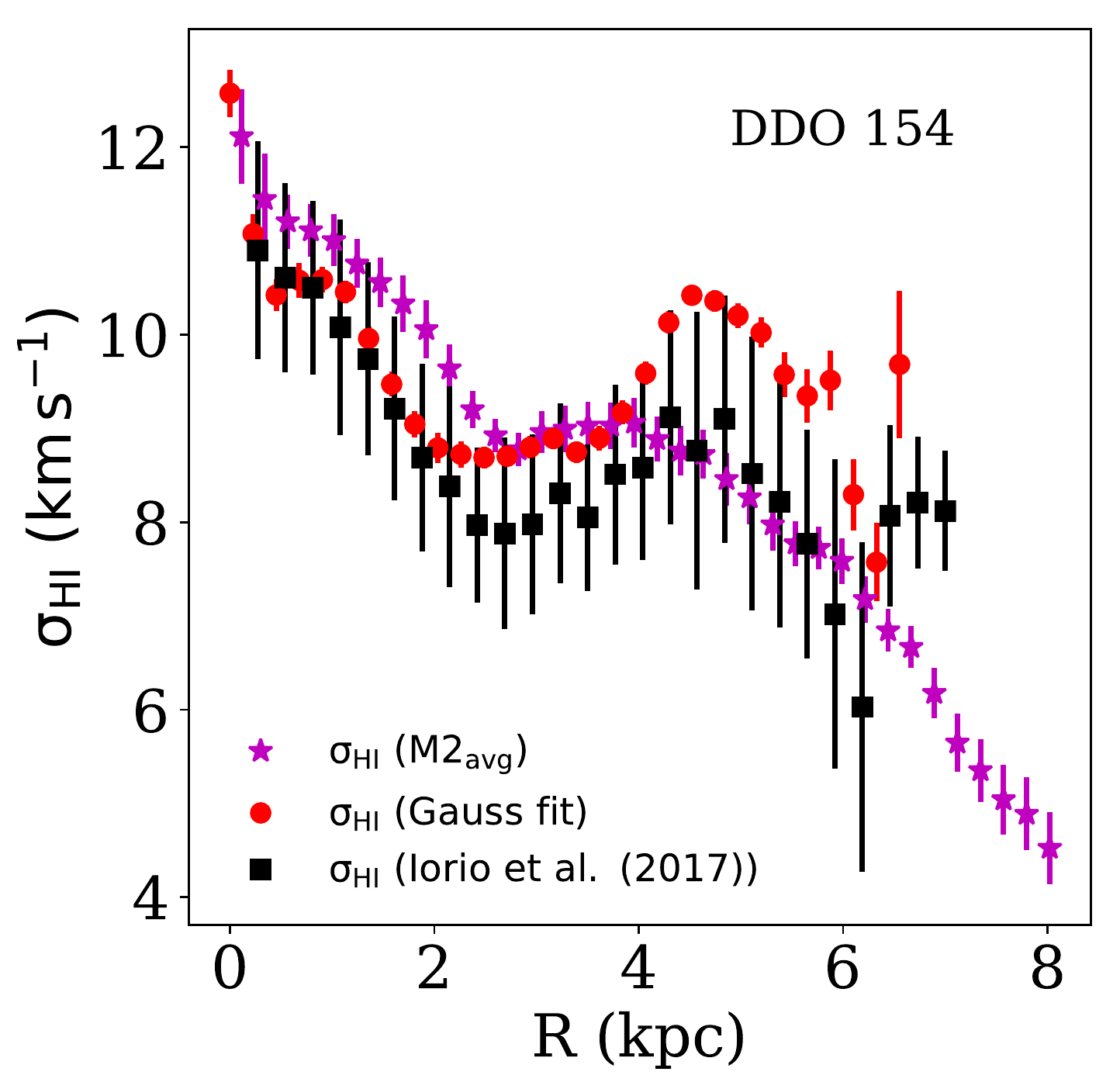}}
\end{tabular}
\end{center}
\caption{Comparison of the $\sigma_{HI}$ profiles for DDO 154 as obtained using three different methods. The solid red circles with error bars represent the $\sigma_{HI}$ as estimated by fitting single-Gaussian functions to the stacked spectra. The magenta asterisks with error bars denote $\sigma_{HI}$ calculated by averaging the observed MOMNT2 within radial annuli. The solid black squares with error bars indicate the $\sigma_{HI}$ derived by performing a 3D tilted ring model fitting to the \HI~spectral cube \citep{iorio17}. See the text for more details.}
\label{sigcomp}
\end{figure}

\begin{figure*}
\begin{center}
\begin{tabular}{c}
\resizebox{\textwidth}{!}{\includegraphics{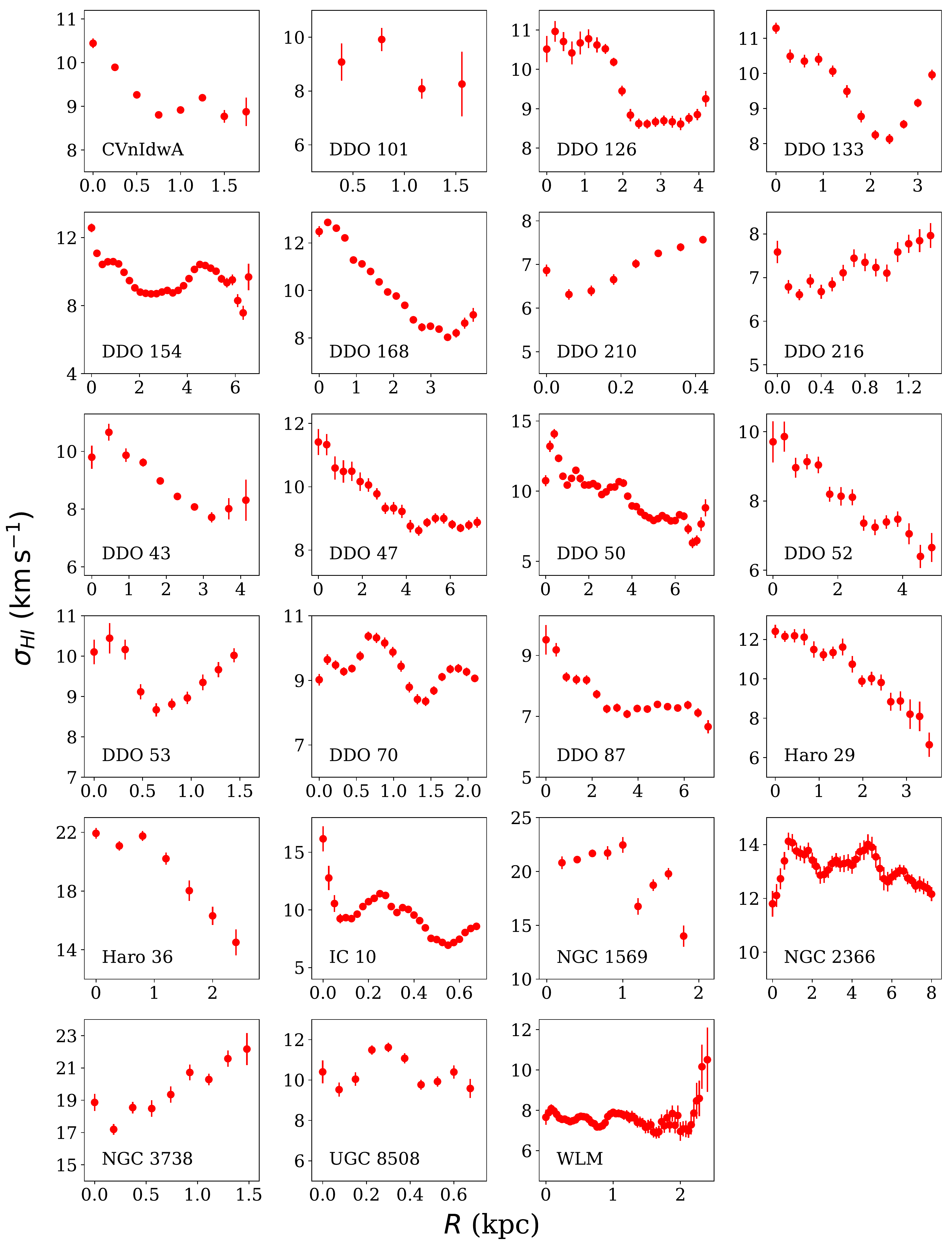}}
\end{tabular}
\end{center}
\caption{The $\sigma_{HI}$ profiles of our sample galaxies. Each panel represents the $\sigma_{HI}$ profile of an individual galaxy, as mentioned in the bottom left corner of the panel.}
\label{sig_all}
\end{figure*}

In Fig.~\ref{sprof}, we show example stacked spectra for the galaxy DDO 154 at four different radii. At a radius of $\sim 1.5$ kpc (the top left panel), $\sim$ 1400 spectra are found to have SNR $\textgreater$ 5. The stacking of these spectra resulted in a stacked spectrum with an SNR of $\sim$ 500, which is much higher than the maximum SNR of any individual spectrum. At a radius of $\sim$ 6 kpc (the bottom right panel), on the other hand, a total of $\sim$ 200 spectra were coadded, resulting in an SNR of $\sim$ 70. As can be seen from the figure, single-Gaussian fits (blue dashed lines) can reasonably describe the spectra and its widths. We emphasize here that at high SNR, the MOMNT2 and the width of the fitted single-Gaussian can represent the $\sigma_{HI}$ equally well. However, at a lower SNR, the MOMNT2 calculation can be affected by the presence of residual noise, especially in the wings (see, e.g., the bottom right panel of Fig.~\ref{sprof}). In such scenarios, the sigma of the single-Gaussian fits to the spectra can better represent the widths than a simple MOMNT2. To estimate the errors in the calculated \HI~velocity dispersion, at each radius, we construct realizations of 100 stacked spectra by boot-strapping individual spectra from the original pool and stack them. We use these 100 stacked spectra and fit them all with single-Gaussian profiles. We then calculate the boot-strapped error as the standard deviation of the widths of these stacked spectra. We add this standard deviation to the fitting uncertainty in quadrature to calculate the total error on our estimated $\sigma_{HI}$ values. We note that as the spectral resolution of the \HI~cubes of our sample galaxies ( $<$ 2.6 \kms) is much smaller than the inferred \HI~line-widths (see, e.g., Fig.~\ref{sig_all}), no correction is needed to account for the under-sampling of the spectra.

To compare how our derived $\sigma_{HI}$ compares with other estimates, we evaluate the $\sigma_{HI}$ profile in a representative galaxy DDO 154 using two other different methods. First, adopting a similar approach as \citet{tamburro09}, we calculate the $\sigma_{HI}$ by averaging the observed MOMNT2 in annular rings, as denoted by the magenta asterisks in Fig.~\ref{sigcomp}. Secondly, we use the derived $\sigma_{HI}$ profile in DDO 154 by \citet{iorio17} through a rigorous 3D tilted ring fitting using the software $^{3D}$Barolo \citep{diteodoro15} (black squares). As can be seen from the figure, in the inner radii, where the SNR is sufficiently high, all the three estimates of $\sigma_{HI}$ match to each other within error bars. However, at larger radii, the average MOMNT2 continues to decline due to a lack of SNR, whereas, the other two estimates remain steady. Since fitting an individual spectrum with a Gaussian Hermite Polynomial requires a minimum SNR of 5, the estimation of $\sigma_{HI}$ at the outermost radii is not possible. However, as the \HI~velocity dispersion in galaxies is found to flatten in the outer discs \citep[see, e.g.,][]{lewis84,das2020}, we assume a constant $\sigma_{HI}$ in these regions which is equal to the value at the last measured point.

In Fig.~\ref{sig_all}, we plot thus obtained $\sigma_{HI}$ profiles for our sample galaxies. As can be seen from the figure, there is a considerable variation in the $\sigma_{HI}$ profiles as a function of radius. It is vital to capture this variation while solving the hydrostatic equilibrium equation as the vertical velocity dispersion can significantly influence the three-dimensional density distribution of the \HI.

\begin{table*}
\caption{Rotation curve fit parameters}
\begin{threeparttable}
\begin{tabular}{lc c c c c c}
\hline
Name  &  Fit type 	  &   $m$    & $c$   &  $V_{max}$   &  $R_{max}$   &   $n$ \\
      & (Linear/Brandt)   &  ($km \thinspace s^{-1} \thinspace kpc^{-1}$) & ($km \thinspace s^{-1}$) & ($km \thinspace s^{-1}$) & kpc &  \\
\hline
CVnIdwA & L & 10$\pm$1 & 2.0$\pm$1.3 & - & - & -\\ 
DDO 101 & B & - & - & 66.1$\pm$2.2 & 3.40$\pm$0.82 & 0.69$\pm$0.13\\ 
DDO 126 & B & - & - & 38.0$\pm$0.4 & 3.61$\pm$0.21 & 3.01$\pm$0.43\\ 
DDO 133 & B & - & - & 46.7$\pm$1.5 & 4.03$\pm$0.71 & 1.43$\pm$0.30\\ 
DDO 154 & B & - & - & 48.5$\pm$0.4 & 7.44$\pm$0.51 & 0.92$\pm$0.06\\ 
DDO 168 & B & - & - & 61.7$\pm$0.6 & 2.60$\pm$0.09 & 5.81$\pm$0.95\\ 
DDO 210 & L & 33$\pm$1 & 1.5$\pm$0.2 & - & - & -\\ 
DDO 216 & B & - & - & 19.4$\pm$8.2 & 3.89$\pm$7.48 & 0.73$\pm$0.82\\ 
DDO 43 & B & - & - & 36.7$\pm$0.7 & 3.33$\pm$0.34 & 2.74$\pm$0.71\\ 
DDO 47 & B & - & - & 64.6$\pm$1.5 & 5.83$\pm$0.30 & 10.40$\pm$4.95\\ 
DDO 50 & B & - & - & 35.3$\pm$0.4 & 5.58$\pm$0.51 & 1.02$\pm$0.14\\ 
DDO 52 & B & - & - & 67.0$\pm$3.5 & 10.24$\pm$2.50 & 0.96$\pm$0.18\\ 
DDO 53 & L & 23$\pm$2 & 1.2$\pm$1.2 & - & - & -\\ 
DDO 70 & L & 17$\pm$1 & 8.7$\pm$1.1 & - & - & -\\ 
DDO 87 & B & - & - & 65.5$\pm$6.2 & 20.21$\pm$7.31 & 0.89$\pm$0.19\\ 
Haro 29 & B & - & - & 34.6$\pm$0.6 & 3.14$\pm$0.68 & 0.55$\pm$0.17\\ 
Haro 36 & L & 14$\pm$1 & 10.6$\pm$1.7 & - & - & -\\ 
IC 10 & B & - & - & 43.2$\pm$18.5 & 1.06$\pm$1.12 & 2.38$\pm$2.49\\ 
NGC 1569 & L & 13$\pm$1 & 12.9$\pm$1.0 & - & - & -\\ 
NGC 2366 & B & - & - & 58.1$\pm$0.2 & 6.52$\pm$0.17 & 1.43$\pm$0.06\\ 
NGC 3738 & B & - & - & 135.0$\pm$10.7 & 2.63$\pm$1.04 & 1.10$\pm$0.37\\ 
UGC 8508 & L & 22$\pm$1 & 4.4$\pm$0.7 & - & - & -\\ 
WLM & B & - & - & 35.8$\pm$0.3 & 3.09$\pm$0.13 & 1.51$\pm$0.08\\ 
\hline
\end{tabular}
\label{tab:rotcur}
\end{threeparttable}
\end{table*}

\begin{figure*}
\begin{center}
\begin{tabular}{c}
\resizebox{1.\textwidth}{!}{\includegraphics{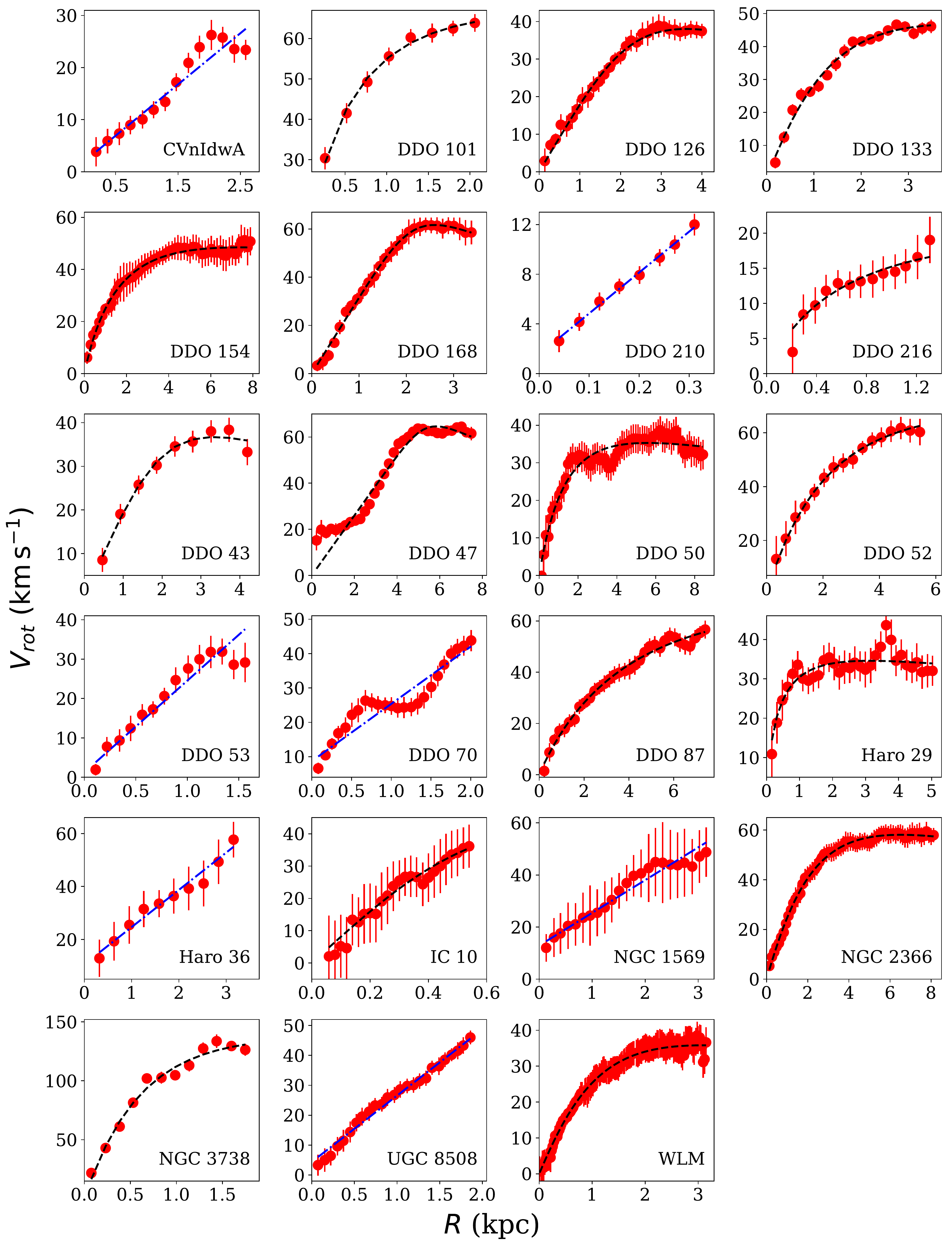}}
\end{tabular}
\end{center}
\caption{The rotation curves of our sample galaxies. Each panel represents the rotation curve of an individual galaxy, as mentioned in the bottom right corner of the panel. The solid red circles with error bars represent the extracted rotation velocities as a function of radius as obtained by a tilted ring fitting \citep{oh15}. The black dashed curves represent a Brandt-profile fit to the rotation curves, whereas the blue dashed-dotted lines represent a straight line fit to the rotation curves. See the text for more details.}
\label{rotcur}
\end{figure*}

The other input to Eq.~\ref{eq5} is the observed rotation curve, which is necessary to compute the radial term (last term on the RHS). The rotation curve of our sample galaxies are extracted by \citet{oh15} using a tilted-ring model fitting to the 2D velocity field. Thus obtained rotation curves are plotted in Fig.~\ref{rotcur}. It should be noted that though the discs of dwarf galaxies show signs of systematic rotation \citep[see, e.g.][]{begum08c}, yet significant irregularities observed in their rotation curves (as can be seen from the figure). In Eq.~\ref{eq5}, the RHS requires a first-order derivative of the rotation curve. This implies any sudden change/jump in the rotation curve (could be due to measurement errors) might lead to an unphysical value of the derivative and hence, might diverge the solutions. To avoid any such problem, we fit the rotation curves with a Brandt profile \citep{brandt60}, which can be given as

\begin{equation}
v_{rot} (R) = \frac{V_{max}\left(R/R_{max} \right)}{\left(1/3 + 2/3 \left(\frac{R}{R_{max}}\right)^n\right)^{3/2n}}
\end{equation}

\noindent where $v_{rot} (R)$ is the rotation velocity, $V_{max}$ is the maximum velocity attained by the rotation curve, $R_{max}$ is the radius at which the $V_{max}$ is observed. The index $n$ determines how fast or slow the rotation curve reaches the $V_{max}$. The black dashed lines in Fig.~\ref{rotcur}, represent a Brandt-profile fit to the rotation curves of our sample galaxies.

Generally, the rotation curves, which increase as a function of radius and attains a constant flat rotation velocity, can be well described by a Brandt profile. However, for dwarf galaxies, the rotation velocity might not always reach the flat velocity (see, e.g., the rotation curve of DDO 210). In these cases, as there is no obvious $V_{max}$ (and hence $R_{max}$), a Brandt profile can not produce a reasonable fit.  In our sample, there are seven galaxies (CVnIdwA, DDO 210, DDO 53, DDO 70, Haro 36, NGC 1569, and UGC 8508) for which a Brandt-profile fitting could not be performed satisfactorily. For these galaxies, we find that a linear fit can describe the data reasonably well. We emphasize that we do not attempt to interpret the physical implications of the shapes of the rotation curves of our sample galaxies. We only approximate the rotation curves with a smooth function, such as a first-order derivative of the same can be computed. Hence, our choice to approximate some of the rotation curves by straight lines would not impact our results.

In Tab.~\ref{tab:rotcur}, we describe the fit parameters of our rotation curves. The first column shows the name of the galaxies. Column (2) represents if the rotation curve is fitted by a Brandt profile or a straight line. Column (3) and (4) represent the slope and the intercept respectively when the rotation curve is fitted with a straight line whereas column (5), (6) and (7) show the fit parameters when the rotation curve is fitted with a Brandt profile. 

The next input parameter is the dark matter halo, which provides a significant amount of gravity in the hydrostatic equilibrium equation. As described earlier, we do not consider the dark matter halo to be a live component in Eq.~\ref{eq5}; rather, it is considered to be fixed as obtained by the mass modeling of our sample galaxies. For this purpose, we use the mass models as described in \citet{oh15} (Table 2). We note that in \citet{oh15}, the mass modeling was done by considering both a pseudo-isothermal profile and an NFW profile for dark matter halo. However, a pseudo-isothermal profile found to describe the dark matter halos of nearby dwarf galaxies better than an NFW profile \citep{moore94,deblok96,deblok97,deblok01,deblok02,weldrake03,spekkens05,
kuziode06,kuziode08,oh11a,oh11b}. For our sample galaxies, we use a pseudo-isothermal profile to describe the dark matter halos, which can be given as

\begin{equation}
\label{eq_iso}
\rho_h(R) = \frac {\rho_0}{1 + \left(\frac{R}{r_s}\right)^2}
\end{equation}

\noindent where $\rho_h (R)$ is the density of the dark matter halo, $\rho_0$ is the characteristic core density, and $r_s$ is the core radius. These two parameters completely describe a spherically symmetric pseudo-isothermal dark matter halo. 

However, as the dwarf galaxies are highly dark matter dominated, the total gravity in their discs are expected to have a substantial contribution from the dark matter halo. In this sense, the structure of the dark matter halo might have a significant influence in deciding the three-dimensional distribution of \HI. Hence, to investigate the same, we also solve Eq.~\ref{eq5} using NFW dark matter profiles in a couple of galaxies. The NFW profile can be given as \citep{navarrofrenkwhite97}

\begin{equation}
\label{nfw}
\rho_h(R) = \frac {\rho_0}{\frac{R}{r_s} \left( 1 + \frac{R}{r_s}\right)^2}
\end{equation}

\noindent where $\rho_0$ is the characteristic density and $r_s$ the scale radius. As we adopted the dark matter halo parameters from \citet{oh15}, we refer the readers to their Table 2 for the values of the same.

\section{Results and discussion}

\begin{figure}
\begin{center}
\begin{tabular}{c}
\resizebox{0.46\textwidth}{!}{\includegraphics{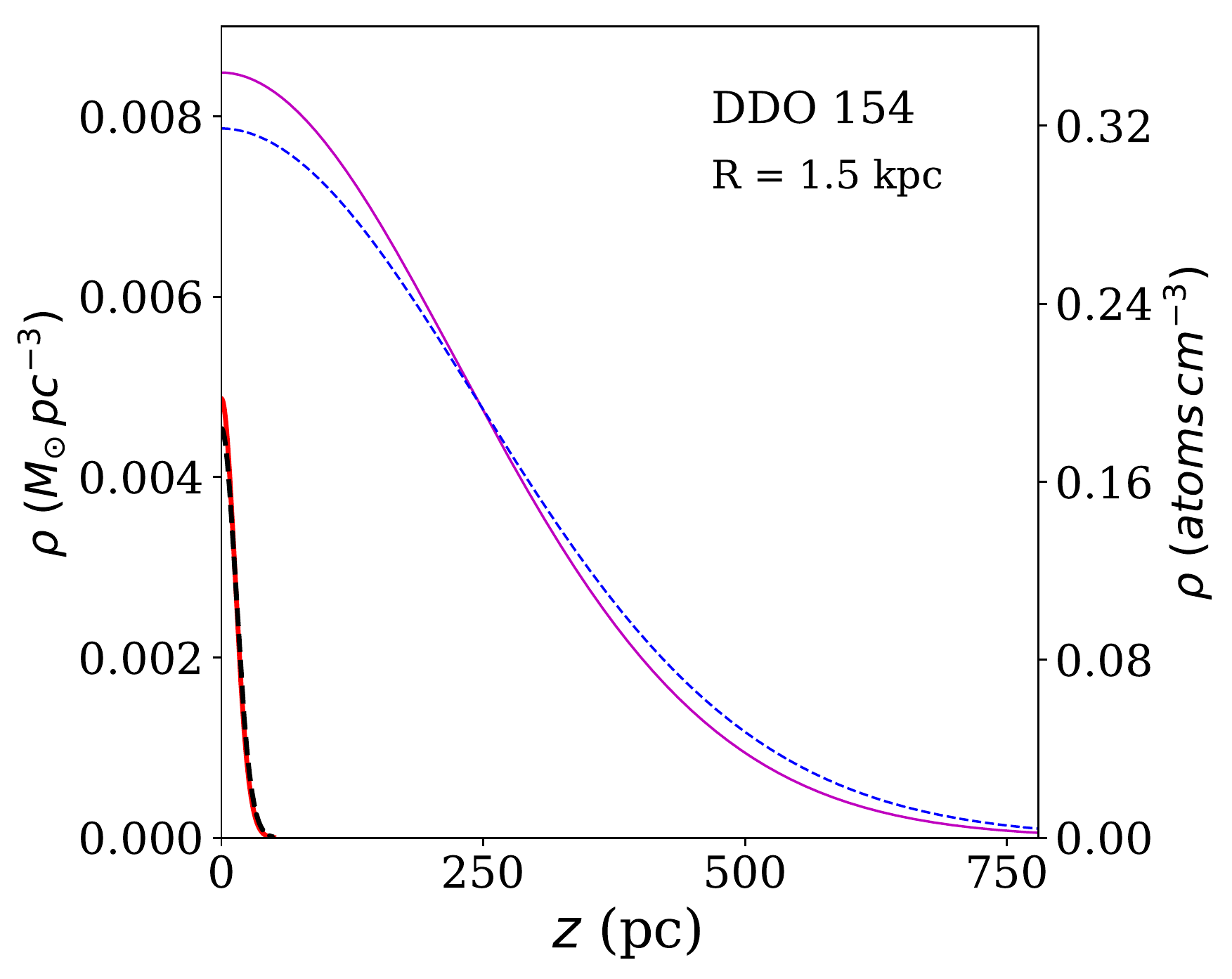}}
\end{tabular}
\end{center}
\caption{\HI~vertical density distributions, derived by solving Eq.~\ref{eq5} for DDO 154 at a radius of 1.5 kpc. The solid magenta and red lines represent the density distribution of the \HI~and the stellar discs, respectively, for an assumed ISO dark matter halo. The dashed blue and the black curves represent the solutions of the \HI~and stars, individually, for an NFW dark matter halo. The NFW halo found to produce lower volume densities close to the midplane (at $R=1.5$ kpc) as compared to the ISO halo. Also, in both cases, the \HI~disc extends to much larger heights as compared to the stellar discs.  See the text for more details.}
\label{solden}
\end{figure}

\begin{figure*}
\begin{center}
\begin{tabular}{c}
\resizebox{1.\textwidth}{!}{\includegraphics{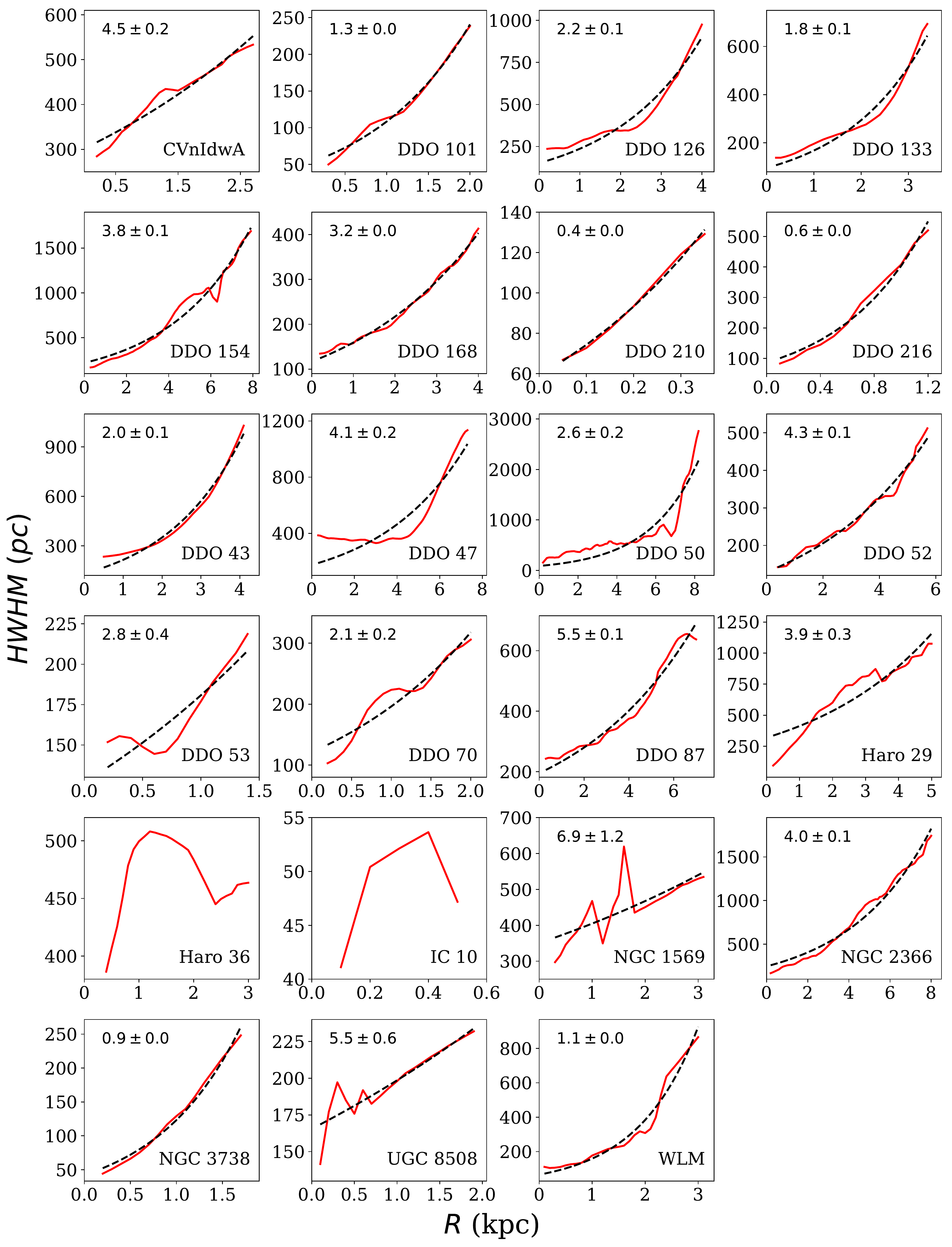}}
\end{tabular}
\end{center}
\caption{The \HI~scale height (HWHM) of our sample galaxies as a function of radius. Each panel represents the HWHM profile for different galaxies, as mentioned in the right bottom corners of the panels. The solid red lines represent the HWHM profiles, whereas the black dashed lines describe an exponential fit to it. The corresponding scale-lengths of the exponential fits are quoted in the top left corners of the respective panels.}
\label{sclh}
\end{figure*}

With the above-mentioned input parameters, we solve Eq.~\ref{eq5} numerically. The details on how we solve the hydrostatic equilibrium equation is presented in the Appendix. We solve Eq.~\ref{eq5} in all our sample galaxies every 100 pc, which is equal to the median spatial resolution of the LITTLE-THINGS galaxies. We emphasize that the star formation rates at the central regions of our sample galaxies are higher than what is observed at their outer discs \citep[see, e.g.,][]{bigiel10b} (even though dwarf galaxies do not have a drastically high star formation rate except a few Blue Compact Dwarfs (e.g., IC 10)). This might lead to the possibility of a non-prevailing hydrostatic equilibrium at these central regions. Moreover, very often, it becomes tough to model the rotation curve at the central regions due to a lack of resolution elements and a possible presence of strong non-circular motions \citep[see, e.g.,][]{iorio17}. Due to these reasons, we exclude a few hundred parsecs region from the center of our sample galaxies while solving the hydrostatic equilibrium equation.

In Fig.~\ref{solden}, we present sample solutions of Eq.~\ref{eq5} for the galaxy DDO 154 at a radius of 1.5 kpc both for the \HI~(the magenta solid and the blue dashed lines) and the stellar (red solid and the black dashed lines) discs. We note that for a single-component isothermal disc stable under its gravity is expected to produce a solution that follows a $sech^2$ law \citep{spitzer42,bahcall84a,bahcall84b}. However, we find that the solutions deviate from a $sech^2$ law due to the coupling between multiple disc components and the dark matter halo. In this case, the solutions are better represented by a Gaussian than a $sech^2$ function. We also find that the \HI~disc in DDO 154 at 1.5 kpc extends to much higher heights ($\gtrsim$ 750 pc) as compared to the stellar disc. An extent of the \HI~disc to a height of $\sim 750$ pc turns out to be much higher than what is observed, for example, in the Milky Way (few hundred parsecs at $R=10$ kpc \citep{kalberla07}). 

For comparison, in Fig.~\ref{solden}, we also plot the solutions assuming an NFW profile for the dark matter halo (the broken lines). As can be seen from the figure, there is a modest effect of the assumed dark matter halo profile on the vertical density distribution. At the central region of DDO 154 (1.5 kpc in this case), an NFW halo provides less amount of gravity close to the midplane as compared to a cored isothermal halo. Subsequently, the NFW halo results in a lower value of the \HI~vertical density around $z=0$ as compared to an ISO halo. However, this does not alter the vertical scale height of the \HI~distribution in a significant way. We discuss this in more detail in subsequent paragraphs.

We use the solutions of Eq.~\ref{eq5} to estimate the \HI~scale heights in our sample galaxies. The \HI~vertical scale height at a radius can be defined as the Half Width at Half Maxima (HWHM) of the \HI~density distribution in the vertical direction. The scale height in galaxies traditionally being used as a measure of the thickness of the baryonic discs. In Fig.~\ref{sclh}, we plot the \HI~scale heights of our sample galaxies as a function of radius. We find that the scale heights in our sample galaxies vary between a few hundred parsecs at the center to $\sim$ a few kiloparsecs at the edge. Not only that, but a large variation in the scale heights are also observed depending on the radius and the galaxy. For example, DDO 52 (12$^{th}$ panel in Fig.~\ref{sclh}) has a scale height of $\sim 500$ pc at the edge of its \HI~disc whereas, DDO 50 (11$^{th}$ panel) found to have a scale height of about $\sim 3$ kpc at its farthest radii. This variation in scale height, in fact, can result in a distribution of shapes of the \HI~discs in dwarf galaxies when observed in projection \citep[see, e.g.,][]{roychowdhury10,roychowdhury13}.

\begin{figure}
\begin{center}
\begin{tabular}{c}
\resizebox{0.46\textwidth}{!}{\includegraphics{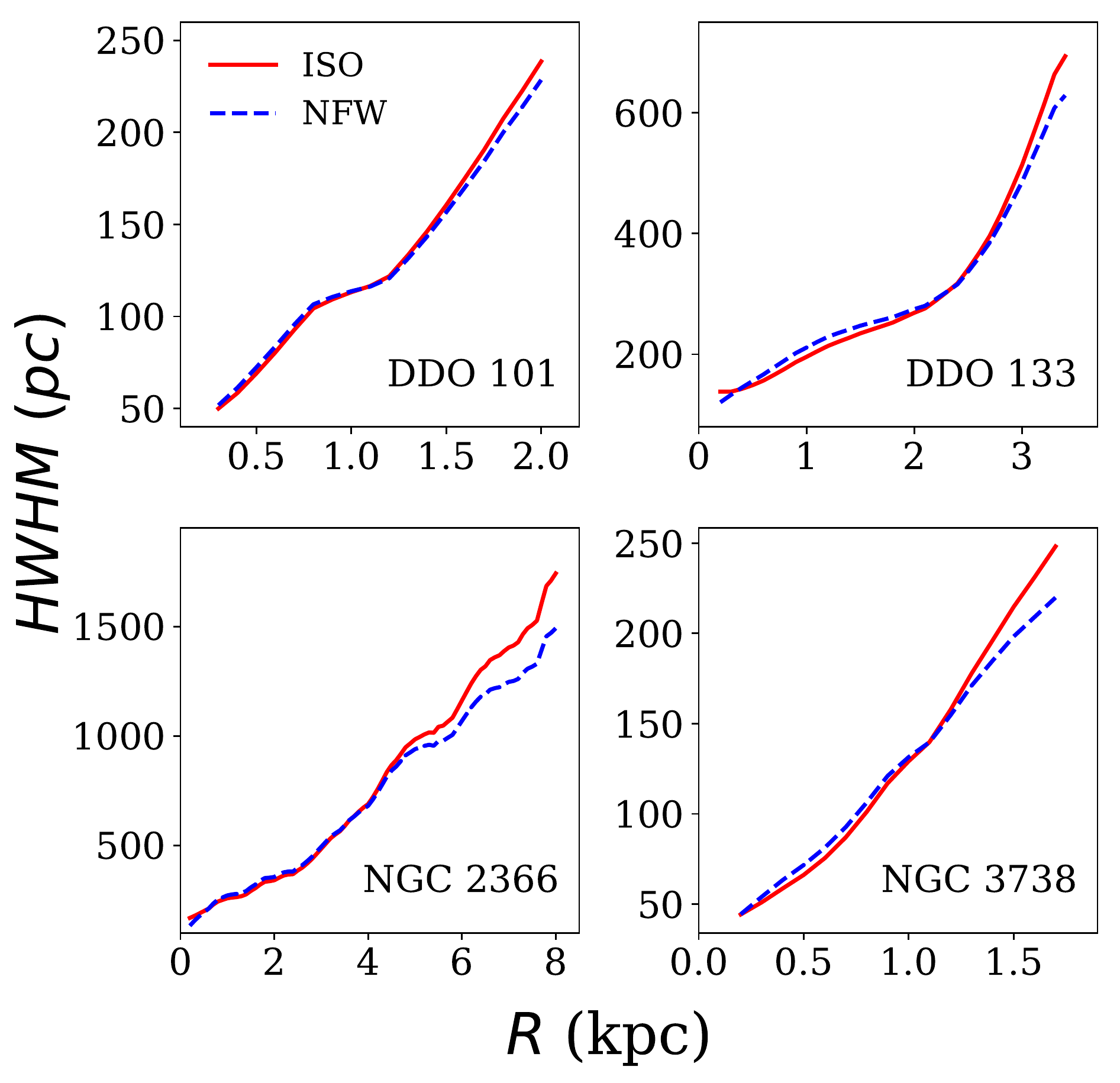}}
\end{tabular}
\end{center}
\caption{The effect of the assumed dark matter distribution on the \HI~scale height profiles. Each panel represents the scale height profile for different galaxies, as quoted at the bottom right corners. The solid red lines represent the HWHM profiles for an assumed ISO dark matter halo whereas, the blue dashed lines show the same for an NFW dark matter halo.}
\label{sclh_dm}
\end{figure}

To investigate the effect of the dark matter halo profile in deciding the \HI~scale height, we solve Eq.~\ref{eq5} for four galaxies from our sample, i.e., DDO 101, DDO 133, NGC 2366 and NGC 3738 using both the ISO and the NFW profiles. Using the density solutions, we estimate the HWHM profiles in these galaxies. In Fig.~\ref{sclh_dm}, we plot these HWHM profiles for both the dark matter halo profiles. We note that NGC 3738 and NGC 2366 are respectively the most and least dark matter dominated systems of our sample. The other two galaxies represent the median population. As can be seen from the figure, different dark matter distributions cause only a marginal difference in the HWHM profiles. Nevertheless, this difference increases slightly at outer radii. It can also be seen from the figure; the ISO halo systematically produces a lower HWHM in the inner galaxy and a higher HWHM in the outer radii. The turn over of the HWHM profile occurs at $\sim 3$ times the core radius of the ISO halo. These results indicate that a different dark matter distribution (ISO or NFW) fails to induce a detectable difference in the vertical \HI~density distribution. However, future sensitive observations with large facilities (e.g., the Square Kilometer Array (SKA)) might be able to distinguish the dark matter distribution in a galaxy by using its vertical \HI~distribution.

\begin{figure}
\begin{center}
\begin{tabular}{c}
\resizebox{0.46\textwidth}{!}{\includegraphics{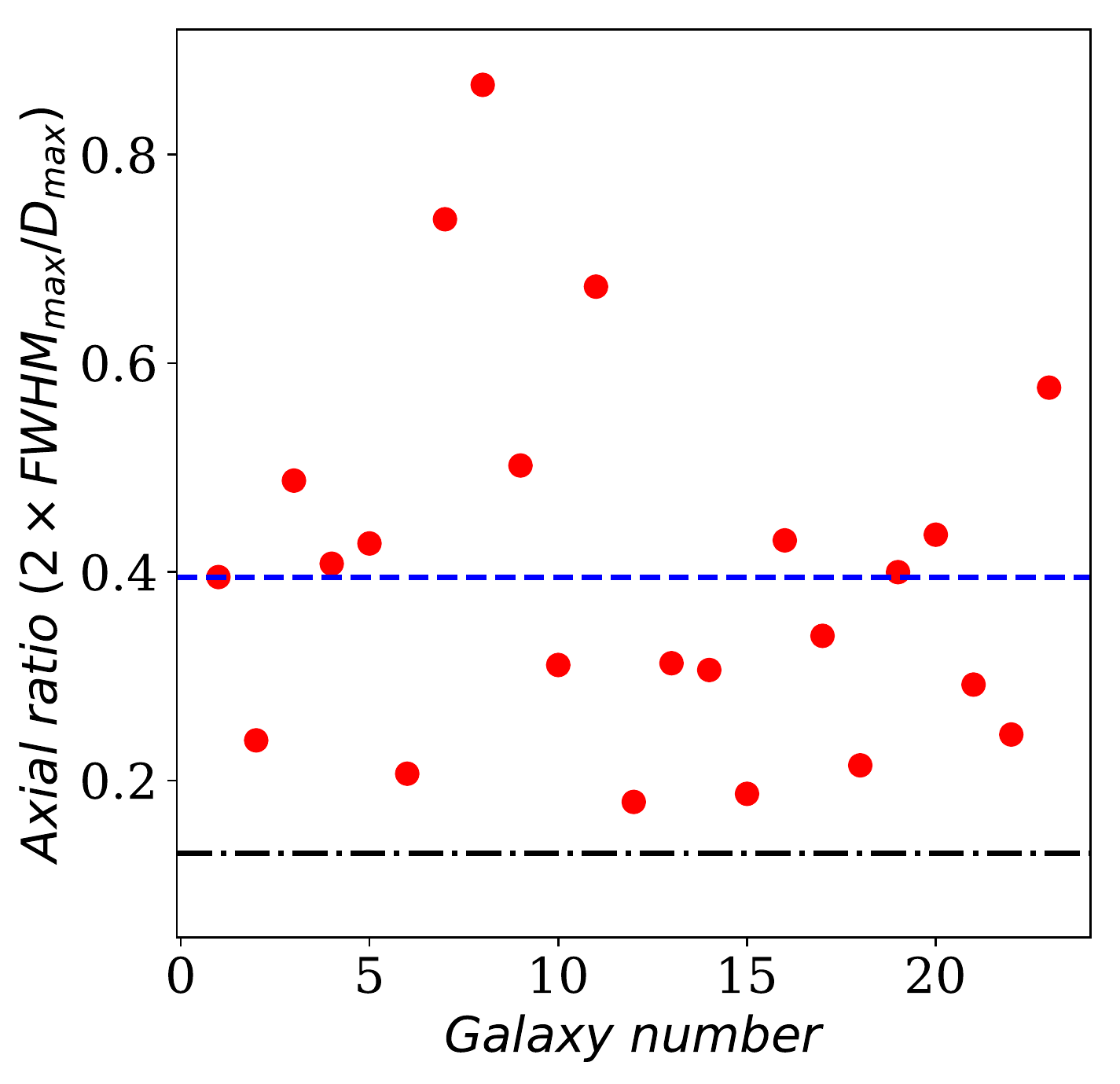}}
\end{tabular}
\end{center}
\caption{Axial ratios of our sample galaxies. The solid red circles are the estimated axial ratios for our sample galaxies. The blue dashed line denotes the median axial ratio of our sample galaxies, which is 0.40. The black dashed line represents the axial ratio for the Milky Way ($\sim 0.13$) at $\sim$ 40 kpc \citep{kalberla07}. This indicates that the axial ratios of dwarf galaxies are much higher than what is observed in a typical spiral galaxy. See the text for more details.}
\label{axial_ratio}
\end{figure}

To further investigate the thickness of the \HI~discs in our sample galaxies, we estimate the axial ratios using the density solutions. The axial ratio can be defined as the ratio of two times the maximum FWHM (a measure of the disc thickness) to the diameter of the \HI~disc. In Fig.~\ref{axial_ratio}, we plot the axial ratios of our sample galaxies (solid red circles). It varies between $\sim 0.18 - 0.87$ with a median value of 0.4 (blue dashed line in the figure). This median axial ratio indicates thick \HI~discs in our sample galaxies. For comparison, we plot the axial ratio for the Milky Way ($\sim 0.13$ at a radius of 40 kpc \citep{kalberla07}), as shown by the black dashed-dotted line. As can be seen from the figure, the axial ratio of a typical dwarf irregular galaxy is much higher than a typical spiral galaxy. This is consistent with observations that dwarf galaxies host a much bloated/puffed \HI~discs \citep{begum08c,roychowdhury13,patra16b}.

\begin{figure}
\begin{center}
\begin{tabular}{c}
\resizebox{0.46\textwidth}{!}{\includegraphics{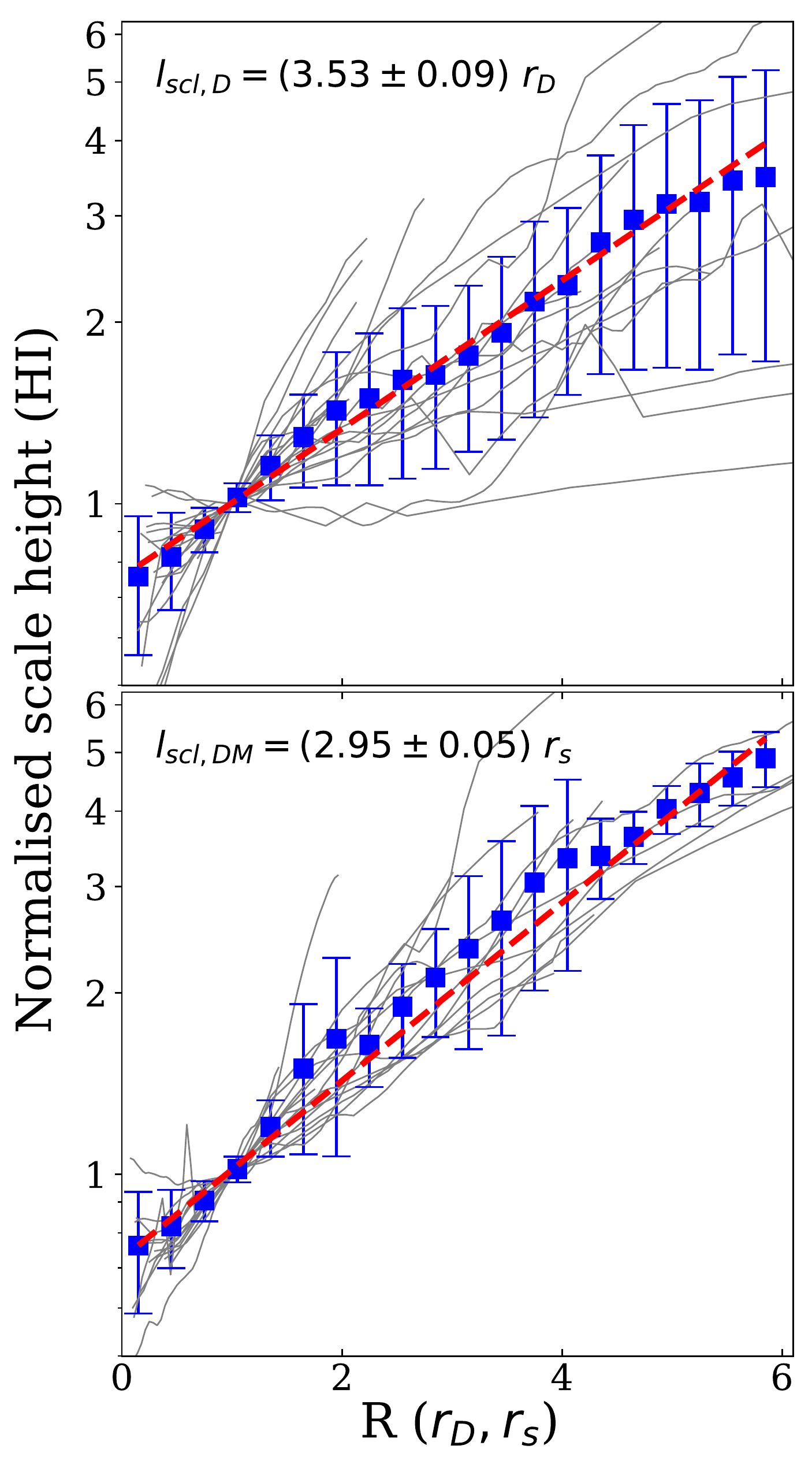}}
\end{tabular}
\end{center}
\caption{Normalized scale height profiles of our sample galaxies. The top panel shows the scale height profiles normalized at $r_D$ (the optical disc scale length), whereas the bottom panel presents the HMHM profiles normalized at $r_s$ (the core radius of the ISO dark matter halo). The grey lines in both the panels represent the normalized \HI~scale height profiles for individual galaxies. The blue squares with error bars denote the average normalized scale heights within a radial bin of 0.3$r_D$ (top panel) or 0.3$r_s$ (bottom panel). The red dashed lines represent exponential fits to the average normalized scale heights. The scale lengths of these exponential fits are quoted at the top left corners of the respective panels.}
\label{norm_scl}
\end{figure}

From Fig.~\ref{sclh}, it can also be seen that the \HI~scale height in our sample galaxies increases as a function of radius, conferring a flaring. However, the nature of the flaring does not seem to be uniform across our sample galaxies. For example, the nature of the \HI~flaring in the galaxy CVnIdwA is quite different from the same in the galaxy DDO 126. To quantify this \HI~flaring, we fit the \HI~scale heights with an exponential function of the form $h_{scl} = h_0 \exp(R/l_{scl})$. Where $l_{scl}$ represents a characteristics scale length depicting how fast or slow the scale height is rising. This could be a useful theoretical representation of the \HI~discs, which then can be used in numerical modeling of galaxies. The respective scale lengths of the exponential fits are quoted in the top left corners of every panel in the units of kpc. The black dashed lines in all the panels represent the exponential fit. However, it should be noted that an exponential function does not well describe all the scale heights in our sample galaxies (see, e.g., scale heights for DDO 47, DDO 53, DDO 70, etc.). For two galaxies, i.e., Haro 36 and IC 10, we do not fit an exponential function to the inferred scale height profile, since the latter does not increase monotonically as a function of radius. The large variation in the scale height profiles could arise due to the irregular nature of our sample galaxies. Unlike the large spiral galaxies where the surface densities follow a well defined exponential disc, the surface densities in dwarf galaxies are patchy and do not always follow an exponential law. Moreover, many dwarf galaxies have very faint stellar discs at the central region, which in turn provide much less gravity (unlike large galaxies), leading to an extra flaring at the central region. For example, in DDO 53, the stellar surface density at the center is very low as compared to the gas surface density (see the 13$^{th}$ panel of Fig.~\ref{sden}) which results in a puffed up \HI~disc at these region leading to a significant deviation of the scale height from an exponential nature. However, it should be mentioned here that, the solution of Eq.~\ref{eq5} depends on many other factors as well, e.g., dark matter halo, rotation curve, \HI~velocity dispersion, etc. Hence, the nature of the flaring should not be attributed solely to the stellar surface density profile, though it might be a significant one. In fact, in a recent study, \citet{bacchini19a} estimated the \HI~scale height profiles in a sample of 12 nearby disc galaxies assuming a prevailing hydrostatic equilibrium. They found that the \HI~scale height profiles in these galaxies very often better modeled by a linear flaring than an exponential one (see their Fig. 4). However, we note that their galaxies have more massive stellar discs as compared to our sample galaxies. Consequently, we find that this leads to an increase in the amount of total gravity on the \HI~disc and produces less dramatic flaring. Therefore, in dwarf galaxies, a sub-dominant baryonic matter can lead to an exponential flaring of the \HI~disc as opposed to what is observed in large spiral galaxies.

To further investigate the universality of the flaring of the \HI~discs, we adopt a similar approach as used in our earlier study of molecular scale height in spiral galaxies \citep{patra18a,patra19b}. We normalize the \HI~scale heights of all our sample galaxies to unity at a radius of $r_D$ (optical disc scale length). In Fig.~\ref{norm_scl} top panel, we plot the normalized scale heights as a function of the normalized radius for all our sample galaxies except DDO 43. DDO 43 has a $r_D = 0.41$ kpc, where we did not solve the hydrostatic equation. From the figure, it can be seen that there is a significant scatter in the normalized scale height profiles as opposed to what is found for spiral galaxies (albeit for molecular gas) \citep[see, e.g.,][]{patra18a,patra19b}. However, the dwarf galaxies are known to be dark matter dominated with a rather sub-dominant baryonic component. This suggests that the nature of the flaring could be regulated by the structure of the dark matter halo more than the stellar disc. Hence, in the bottom panel of Fig.~\ref{norm_scl}, we plot the \HI~scale heights of our sample galaxies normalized to the core radius ($r_s$) of the ISO dark matter halo. In both the panels, we exclude Haro 36 and IC 10 as we do not choose to describe their scale height profiles by an exponential function. As can be seen from this panel, the \HI~scale heights normalized by the core radius show much less scatter than what is observed in the top panel. This indicates that in dwarf galaxies, the dark matter distribution mainly decides the nature of flaring in the \HI~discs. We further bin the normalized radius with a bin size of 0.3$r_D$ or 0.3$r_s$ and compute the average value of the normalized scale height in every bin (solid blue squares with error bars in both the panels). To obtain a general relation between the normalized scale height and the normalized radius, we fit these mean values with an exponential function $h_{scl} = h_0 \exp{(R/l_{scl})}$. We find the scale lengths as $l_{scl,D} = (3.53 \pm 0.09) \ r_D$ and $l_{scl,DM} = (2.95 \pm 0.05) \ r_s$ for our galaxies. However, despite normalizing by core radius, the \HI~scale height profiles still show significantly higher scatter than what is observed for spiral galaxies. This suggests that the structures of the \HI~discs in dwarf galaxies are not as universal as it could be for spiral galaxies.

\section{Conclusion}

We model the baryonic discs in dwarf galaxies as two-component systems consist of stars and atomic hydrogen in vertical hydrostatic equilibrium under their mutual gravity in the external force field of the dark matter halo. We subsequently set up the joint Poisson's-Boltzmann equation of hydrostatic equilibrium. We solve the hydrostatic equilibrium equation numerically using an 8$^{th}$ order Runge-Kutta method as implemented in the python package `{\tt scipy}'. The solutions provide a detailed three-dimensional distribution of the \HI~in a galaxy.

Further, the hydrostatic equilibrium equation is solved in a sample of 23 local volume dwarf galaxies from the LITTLE-THINGS survey. This is the largest sample to date for which detailed hydrostatic modeling of the \HI~discs is performed. The vertical velocity dispersion of the \HI, $\sigma_{HI}$, is one of the crucial inputs required to solve the hydrostatic equilibrium equation. As this can influence the vertical distribution of the \HI~directly, a precise measurement of the same is essential. We stack the line-of-sight \HI~spectra from the LITTLE-THINGS survey data in radial bins to produce high SNR stacked spectra for individual galaxies. These stacked spectra are then fitted with single-Gaussian profiles to estimate the \HI~velocity dispersion as a function of radius. Using this $\sigma_{HI}$ profile, and other input parameters, we solve the Poisson's-Boltzmann equation in our sample galaxies every 100 pc and produce a detailed three-dimensional \HI~density distribution in them. We further investigate the effect of the dark matter halo profile on the vertical density distribution. We find that the choice of the dark matter density distribution (ISO or NFW) does not imprint a detectable difference in the \HI~density distribution. Highly sensitive observations with large facilities would be essential to distinguish a dark matter halo type in a galaxy using its vertical \HI~distribution.

We further use the density solutions to estimate the vertical \HI~scale height as a function of radius. We find that the \HI~in dwarf galaxies flares with an increasing radius. The flaring found in our galaxies is significantly higher as compared to what is observed in typical spiral galaxies. To quantify the flaring, we fit the scale heights with an exponential function. It is found that by an large, an exponential function describes the flaring reasonably well. However though for a few galaxies, an exponential function does not provide an excellent fit. This indicates a considerable variation in the characteristic structure of the \HI~discs in dwarf galaxies.

We further test the universality of this \HI~flaring by normalizing the scale height profiles to unity at both their optical disc scale length and the core radius of the dark matter halo. Unlike spiral galaxies, which show a tight exponential behavior, dwarf galaxies are found to have a significant scatter in their normalized scale height profiles when normalized by the scale lengths of the optical disc. This scatter significantly reduces when the normalization is done using the core radius. This indicates that the structure of the dark matter halo in dwarf galaxies predominantly decides the nature of the flaring in \HI~discs. An exponential fit to the average normalized scale height profile (by the core radius, $r_s$) results in a normalized scale length $l_{scl,DM} = (2.95 \pm 0.05) \ r_s$.

To investigate how the thickness of the \HI~discs in dwarf galaxies compare to that of the spiral galaxies, we use the scale height profiles of our sample galaxies to estimate the axial ratios defined as the ratio of twice the maximum FWHM to the diameter of the \HI~disc. The axial ratio in our sample galaxies found to vary between $\sim 0.18 - 0.87$ with a median value of 0.40. This median axial ratio is much higher than what is observed in the Milky Way ($\sim 0.13$). This implies that much thicker \HI~discs naturally occurs in dwarf irregular galaxies under vertical hydrostatic equilibrium.

\section{Appendix}
\subsection{Solving the hydrostatic equation}

With the input parameters mentioned in \S~\ref{inputs}, Eq.~\ref{eq5} can be solved in principle. However, this equation cannot be solved analytically even for a minimum number of disc component two. Instead, we solve this equation numerically using an Eight-order Runge-Kutta method as implemented in the python package {\tt `scipy'}.

A detailed description of the approach we adopt to solve Eq.~\ref{eq5} can be found in many earlier studies \citep[see for example][]{narayan05c,banerjee08,banerjee11,patra14,patra18a,patra19b}. Here we give a brief account of the scheme we use to solve the coupled hydrostatic equilibrium equation. 

As Eq.~\ref{eq5} is a second-order partial differential equation, one needs at least two initial conditions to solve it. The two conditions are 

\begin{equation}
\left( \rho_i \right)_{z = 0} = \rho_{i,0} \ \ \ \ {\rm and} \ \ \ \left(\frac{d \rho_i}{dz}\right)_{z=0} = 0
\label{init_cond}
\end{equation}

The second condition originates by the fact that at the mid-plane, the density of both the stellar and the gas discs will be maximum, and hence, the first derivative of the same will be zero. To satisfy the first condition, we presume that the density at the mid-plane (i.e., at $z=0$) is known ($\rho_{i,0}$). However, this mid-plane density is not a directly measurable quantity. Instead, we use the observed surface density to estimate the mid-plane density, $\rho_{i,0}$ indirectly. For any disc component (stars or gas), we start with a trial mid-plane density, $\rho_{i,t}$ and solve Eq.~\ref{eq5} to produce a trial density distribution $\rho_i(z)$. Now this $\rho_i(z)$ can be integrated to calculate the trial surface density, i.e., $\Sigma_i = 2 \int_0^\infty \rho_i(z) dz$. This $\Sigma_i$ then can be compared with the observed surface density to update the trial mid-plane density $\rho_{i,t}$ in the next iteration. We continue this iterative process until the trial $\Sigma_i$ matches the observed value with better than 1\% accuracy.

Here we would like to mention that, in all previous studies, updating the trial mid-plane density was done manually. This might necessitate a large effort to solve Eq.~\ref{eq5} in a comprehensive number of radial points for a significant number of sample galaxies. To overcome this shortcoming, we automatize the selection of $\rho_{i,0}$ using a bisection method. For every radius, we select two trial mid-plane densities such that they produce trial surface densities which enclose the observed value. Then in every iteration using a bisection approach, we narrow down the range of the trial mid-plane density, such as it produces a trial surface density closer to the observed one. We continue this bisection process until we find a $\rho_{i,0}$, producing an observed $\Sigma_i$ with better than 1\% accuracy. We note that for our sample galaxies, the bisection method converges quickly within 10s of iterations without requiring any manual input.

Further, Eq.~\ref{eq5} represents two partial differential equations in $\rho_s$ and $\rho_{HI}$, which are coupled through gravity, the first term on the RHS. Ideally, both these equations should be solved simultaneously, which is not possible. Instead, we adopt an iterative approach to include a constant density distribution (hence gravity) of one component while solving for the other one. In the first iteration, both the stars and gas are solved separately, considering no coupling. This means the density of gas made to be zero while solving for stars and vice versa. In the next iterations, one component included in the equation as a constant term while solving the other one. For example, in the first iteration, we solve for stars and gas with no coupling and produce $\rho_{s,1}(z)$ and $\rho_{HI,1}(z)$. Here the second index indicates the iteration number. In the second iteration, while solving for stars, we introduce $\rho_{HI,1}(z)$ in the first term of RHS in Eq.~\ref{eq5} and keep it constant. Now in the second iteration, the density distribution of the stars is produced in the presence of a gas disc, albeit the gas density distribution is not entirely correct. Similarly, we solve for gas in the presence of star and produce $\rho_{s,2}(z)$ and $\rho_{HI,2}(z)$, which are slightly better than the density distribution obtained in the first iteration. We continue this iterative method until the density distributions of the disc components converge and do not change at better than 1\% level. We find that for our sample galaxies, this method quickly converges within a few iterations.

It should be emphasized here that Eq.~\ref{eq5} has to be solved for a large number of radial points for all our sample galaxies. As solving Eq.~\ref{eq5}, at a radius, does not depend on the solutions at any other radius, it can be solved parallelly. We employ MPI base parallel coding to solve Eq.~\ref{eq5} simultaneously at multiple radii. This significantly reduces the computation time to generate the density solutions. It should be noted that a manual inspection to search for the correct trial mid-plane density, $\rho_{i,t}$ would have been restricted the implementation of parallel processing. In that sense, the bisection approach plays an indispensable role in the automation of the differential equation solver.

\bibliographystyle{mn2e}
\bibliography{bibliography}

\end{document}